\documentclass{article}
\usepackage{times}

\usepackage{microtype}
\usepackage{graphicx}
\usepackage{booktabs} 
\usepackage{makecell}
\usepackage{tabularx}
\usepackage{url}

\usepackage[T1]{fontenc}

\usepackage{float}

\usepackage{stfloats}

\usepackage{listings}

\usepackage{hyperref}
\usepackage{xurl}

\usepackage[most]{tcolorbox}
\usepackage{amsmath}
\usepackage{amssymb}
\usepackage{mathtools}
\usepackage{xcolor}

\usepackage{color}
\usepackage{colortbl}

\usepackage{eso-pic}
\usepackage{lmodern}
\usepackage{setspace}
\usepackage{fancyhdr}
\usepackage{caption}

\usepackage{nebius}

\IfFileExists{latexml.sty}{\usepackage{latexml}}{\newif\iflatexml}

\newcommand{\firstauthor}{Vladimir Dobrovolskii}
\newcommand{\secondauthor}{Ibragim Badertdinov}
\newcommand{\thirdauthor}{Alexander Golubev}
\newcommand{\paperaffiliation}{Nebius}
\newcommand{\correspondingemail}{alex\_golubev@nebius.com}

\title{Longtail-SWE: High-Quality SWE Tasks for Underrepresented Languages}
\author{%
  \firstauthor\\\paperaffiliation
  \and \secondauthor\\\paperaffiliation
  \and \thirdauthor\footnote{Correspondence to: \correspondingemail\\\mbox{}}\\\paperaffiliation
}
\date{}

\makeatletter
\hypersetup{
  pdftitle={\@title},
  pdfauthor={\firstauthor, \secondauthor, \thirdauthor}
}
\makeatother

\renewenvironment{abstract}{\vskip.075in\centerline{\large\bfseries
Abstract}\vspace{0.5ex}\begin{quote}}{\par\end{quote}\vskip 1ex}

\renewcommand{\headrulewidth}{1pt}

\AddToShipoutPicture*{
  \AtTextUpperLeft{
    \raisebox{5pt}{
      \includegraphics[width=80px]{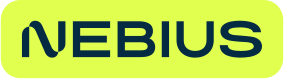}
    }
  }
}

\definecolor{bgcolor}{HTML}{EFF4F9}

\newtcolorbox{roundedbox}{
  enhanced,
  colback=bgcolor,
  frame hidden,
  arc=3mm,
  auto outer arc,
  left=4pt, right=4pt,
  top=6pt, bottom=6pt,
}

\iflatexml\else
\makeatletter
\renewcommand{\@maketitle}{%
\begin{roundedbox}
  \centering

  \vspace{0.3em}
  {\fontsize{18pt}{22pt}\selectfont\bfseries
    \@title
  }\par
  \vspace{1.2em}

  \begin{spacing}{1.15}
  \textbf{\firstauthor}\textsuperscript{1},\ %
  \textbf{\secondauthor}\textsuperscript{1},\ %
  \textbf{\thirdauthor}\textsuperscript{1*}
  \end{spacing}
  \vspace{0.5em}

  {\small
  \textsuperscript{1}\paperaffiliation\\
  \vspace{0.3em}
  \textsuperscript{*}Correspondence to: \correspondingemail\\
  }
  \vspace{0.8em}

  \begin{abstract}
Challenging repository-level benchmarks remain limited for many programming language ecosystems, where filtering scarce candidates with ambiguous descriptions or defective tests can discard valuable engineering problems. We introduce Longtail-SWE, a benchmark of 202 issue-resolution tasks across 13 languages with limited coverage in existing repository-level benchmarks: Clojure, Dart, Elixir, Erlang, F\#, Haskell, Julia, Kotlin, Lua, OCaml, R, Scala, and Swift. Its construction uses a solver--judge--repairer pipeline: a solver attempts each candidate task, and a judge reviews the task together with the solver's trajectory and test results to guide revisions to tests and task descriptions. These revisions aim to correct tests that reject valid alternative implementations or accept incorrect solutions while retaining challenging tasks and preserving their intended requirements. We evaluate nine model configurations, including frontier models, and analyze task-solving performance, consistency across repeated attempts, and resource usage. We release all benchmark instances, executable environments, complete construction and evaluation tooling, and agent trajectories, together with an additional set of easier tasks to support evaluation of smaller models.
\end{abstract}

\end{roundedbox}
}
\makeatother
\fi

\begin{document}
\onecolumn

\makeatletter
\setlength{\@fptop}{0pt}
\setlength{\@dblfptop}{0pt}
\makeatother

\maketitle
\thispagestyle{empty}
% The PDF title box already includes the abstract.
\iflatexml
  \begin{abstract}
Challenging repository-level benchmarks remain limited for many programming language ecosystems, where filtering scarce candidates with ambiguous descriptions or defective tests can discard valuable engineering problems. We introduce Longtail-SWE, a benchmark of 202 issue-resolution tasks across 13 languages with limited coverage in existing repository-level benchmarks: Clojure, Dart, Elixir, Erlang, F\#, Haskell, Julia, Kotlin, Lua, OCaml, R, Scala, and Swift. Its construction uses a solver--judge--repairer pipeline: a solver attempts each candidate task, and a judge reviews the task together with the solver's trajectory and test results to guide revisions to tests and task descriptions. These revisions aim to correct tests that reject valid alternative implementations or accept incorrect solutions while retaining challenging tasks and preserving their intended requirements. We evaluate nine model configurations, including frontier models, and analyze task-solving performance, consistency across repeated attempts, and resource usage. We release all benchmark instances, executable environments, complete construction and evaluation tooling, and agent trajectories, together with an additional set of easier tasks to support evaluation of smaller models.
\end{abstract}

\fi

\section{Introduction}
\label{sec:introduction}

Repository-level benchmarks evaluate whether coding agents can resolve real software issues in existing codebases~\cite{swebench}. SWE-bench~\cite{swebench} has introduced a pipeline for task collection for Python, and later Multi-SWE-bench~\cite{multiswebench}, SWE-PolyBench~\cite{swepolybench}, and SWE-Bench Pro~\cite{swebenchpro} have broadened this setting beyond it. However, many programming language ecosystems remain sparsely represented in dedicated issue-resolution evaluations. This limits the evidence available about how well agents work with different libraries, toolchains, and development practices. Extending this coverage requires executable tasks that are both challenging and reliably evaluated.

Such tasks are commonly built from historical issues and pull requests (PRs), whose original fixes serve as reference patches. However, checking that a reference patch resolves test failures without regressions does not establish evaluation reliability~\cite{sweabs}. PR-specific tests may reject valid alternative implementations, and descriptions may omit requirements that agents cannot infer from the available context~\cite{swebenchverified}, while insufficient tests may accept incorrect solutions~\cite{utboost}.

Filtering can discard repairable tasks and bias benchmark composition: the SWE-bench Multilingual release notes that requiring clear descriptions and unambiguous tests favors small code changes~\cite{swebenchmultilingual}. Recovering such candidates is especially important in underrepresented ecosystems with limited task pools. Building on prior work on test strengthening~\cite{sweabs}, specification repair~\cite{swebenchplusplus}, and agent-informed review~\cite{swemarathon}, we jointly refine descriptions and tests to retain challenging tasks.

We introduce \textbf{Longtail-SWE}, a benchmark of 202 tasks derived from real pull requests across 13 programming languages: Clojure, Dart, Elixir, Erlang, F\#, Haskell, Julia, Kotlin, Lua, OCaml, R, Scala, and Swift. These languages span diverse programming paradigms and application domains, providing a dedicated setting for examining agent performance beyond the ecosystems emphasized by existing issue-resolution benchmarks.

We construct Longtail-SWE using a solver--judge--repairer pipeline. A solver attempts each task; a judge compares its trajectory and test results with the task description, reference patch, and tests to distinguish solver errors from task defects. The judge's findings guide a repair agent in revising tests and appending clarifications to the description. These revisions aim to reject incorrect solutions without excluding valid alternatives or prescribing implementation details, while preserving the task's intended engineering requirements.

Our contributions are threefold:

\begin{itemize}
    \item \textbf{Task recovery and refinement.} Our central contribution is a solver--judge--repairer pipeline that coordinates revisions to task descriptions and tests, allowing scarce repository-level tasks with repairable defects to be retained rather than discarded.
    \item \textbf{Benchmark and companion set.} Applying this method yields Longtail-SWE, a benchmark of 202 challenging tasks across 13 underrepresented programming languages, together with a companion set of easier tasks.
    \item \textbf{Empirical analysis.} We quantify how task recovery and refinement improve executable candidate yield and final task retention. We evaluate nine model configurations, comparing task-solving performance across languages, consistency across repeated attempts, and resource usage.
\end{itemize}

\section{Related Work}
\label{sec:related}

\paragraph{Repository-level and multilingual benchmarks.}
SWE-bench~\cite{swebench} established repository-level issue-resolution evaluation in Python, and Multi-SWE-bench~\cite{multiswebench}, SWE-bench Multilingual~\cite{swebenchmultilingual}, and SWE-PolyBench~\cite{swepolybench} extended it to other languages. Specialized coverage includes Kotlin repositories in the Kotlin Benchmark~\cite{kotlinbenchmark} and Kotlin/Dart mobile applications in MobileDev-Bench~\cite{mobiledevbench}. Longtail-SWE complements these efforts with tasks from underrepresented language ecosystems.

\paragraph{Challenging software engineering tasks.}
SWE-Bench Pro~\cite{swebenchpro} targets demanding issue resolution, while SWE-EVO~\cite{sweevo}, SWE-Bench ProMax~\cite{swebenchpromax}, and SWE Refactor Bench~\cite{swerefactorbench} address software evolution, multilingual refactoring, and stack migration, respectively. DeepSWE~\cite{deepswe}, FrontierSWE~\cite{frontierswev2}, and SWE-Marathon~\cite{swemarathon} evaluate long-horizon engineering or research tasks. ProgramBench~\cite{programbench} evaluates program reconstruction with behavioral tests that permit alternative implementations.

\paragraph{Automated task collection and environment construction.}
SWE-rebench~\cite{swerebench} automates task collection and validation from real pull requests. SWE-rebench V2~\cite{swerebenchv2} scales collection to 20 languages and adds diagnostic annotations for task quality. SWE-Bench++~\cite{swebenchplusplus} combines environment construction with test extraction and quality checks. SWE-Factory~\cite{swefactory} uses agents for environment setup, while SWE-Universe~\cite{sweuniverse} validates verifiers on buggy and fixed states and screens for evaluation shortcuts. R2E-Gym~\cite{r2egym} generates tests and descriptions from commits; SWE-smith~\cite{swesmith} synthesizes bugs while reusing repository environments and tests. Terminal-Universe~\cite{terminaluniverse} reconstructs workspaces from agent trajectories to synthesize tasks. Longtail-SWE complements automated task collection with coordinated refinement of descriptions and tests, aiming to retain challenging candidates from scarce task pools in underrepresented language ecosystems.

\paragraph{Task auditing and specification alignment.}
SWE-bench Verified~\cite{swebenchverified} uses expert review to exclude tasks with underspecified issues or unsuitable tests. SPICE~\cite{spice} automates annotations of issue clarity, test adequacy, and expected effort; PAIChecker~\cite{paichecker} detects issue--PR mismatches. Auto Benchmark Audit~\cite{aba} investigates defects in specifications, environments, and grading, including evidence from agent trajectories. SWE-Bench++ supplements underspecified descriptions with interface information from reference patches and filters instances whose failed solver runs reveal task defects. SWE-Marathon~\cite{swemarathon} uses pilot trajectories to distinguish agent failures from task defects, revising and revalidating affected candidates. The Verification Horizon~\cite{verificationhorizon} examines instruction clarity and alignment between instructions and tests as requirements for faithful reinforcement-learning rewards.

\paragraph{Test adequacy and repair.}
Test-suite overfitting is a longstanding concern in automated program repair~\cite{qipatchcorrectness}. EvalPlus~\cite{evalplus} augments functional tests to expose errors in generated programs. For generated postconditions, nl2postcond~\cite{nl2postcond} evaluates both test-set correctness on a reference implementation (\texttt{accept@k}) and bug completeness against incorrect mutants, reflecting the goals of accepting valid solutions and rejecting invalid ones, respectively. UTBoost~\cite{utboost} generates additional tests for SWE-bench to identify incorrect agent patches that passed the original suites. SWE-ABS~\cite{sweabs} combines coverage-guided generation with mutation analysis, relaxing tests that reject semantically equivalent mutants and adding tests against incorrect mutants. SWE-Mutation~\cite{swemutation} evaluates test discrimination using semantic mutants, while XREPOTEST~\cite{xrepotest} benchmarks multilingual repository-level unit test generation. Prior work on reward hackability in code RL environments~\cite{rewardhackabilityaudit} combines reward-hackability auditing with test augmentation checked against reference solutions. Joint curation also appears in SWE-Bench ProMax~\cite{swebenchpromax}, which revises descriptions with expert involvement and removes tests enforcing incidental implementation details or unstated requirements. Longtail-SWE uses solver trajectories and judge feedback to guide coordinated repairs to descriptions and tests, aiming to preserve intended requirements while addressing both incorrect acceptance and incorrect rejection of agent solutions.
\section{Method}
\label{sec:method}

Longtail-SWE builds on SWE-rebench V2 for task collection and executable environment construction. We modify the candidate selection and environment installation procedures, and add a solver--judge--repairer stage that uses solution attempts to identify and repair task defects, followed by benchmark finalization based on additional solver runs for quality and difficulty assessment. The goal is to retain challenging tasks whose descriptions and tests support reliable evaluation.

\subsection{Task Definition and Pipeline Overview}
\label{sec:method-overview}

Longtail-SWE follows the repository-level issue-resolution setting of SWE-bench. Each task consists of a public problem statement, a repository at a pre-fix revision (the \emph{base commit}), an executable environment, a \emph{reference patch}, and an \emph{evaluation test patch}. We obtain the problem statement from a GitHub issue and extract the reference and test patches from its associated merged pull request (PR).

\begin{figure*}[t]
    \centering
    
    \includegraphics[width=\linewidth]{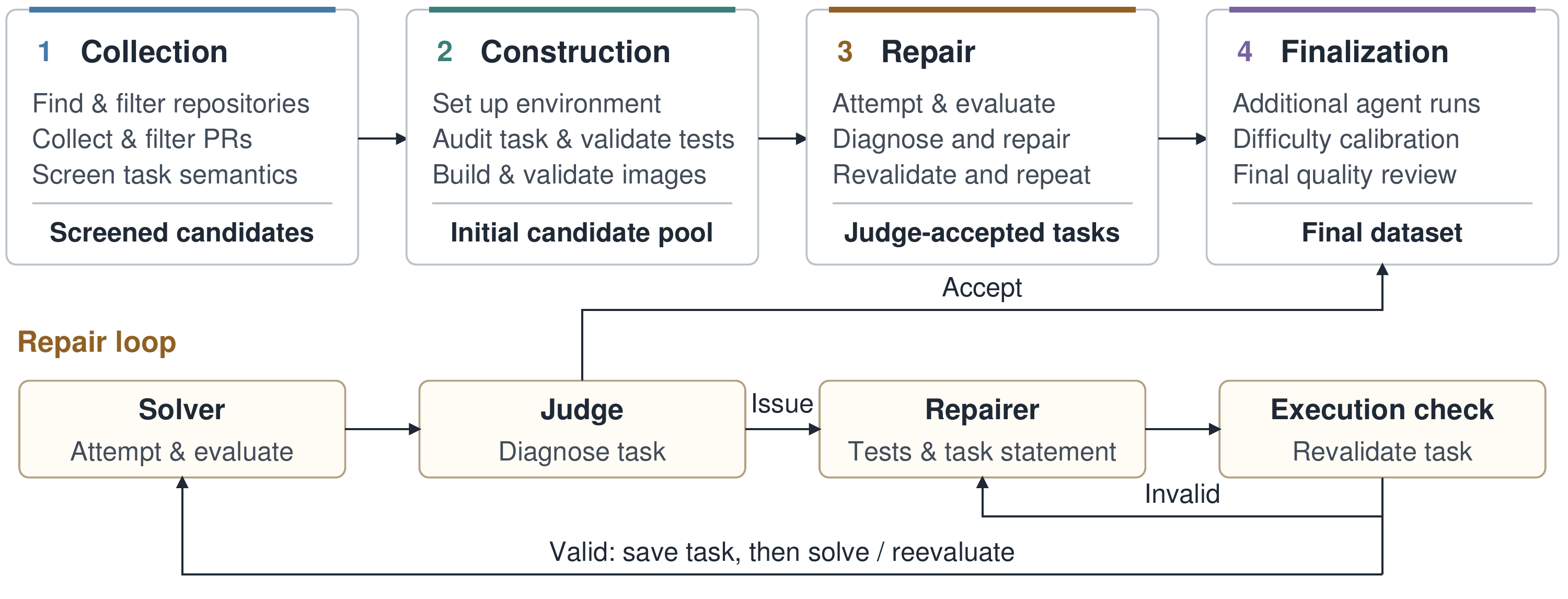}
    \caption{Longtail-SWE pipeline. Executable candidates enter a solver--judge--repairer loop with execution revalidation; judge-accepted tasks proceed to difficulty selection and final quality review.}
    \label{fig:longtail-pipeline}
\end{figure*}

Our pipeline comprises four stages: collection, construction, repair, and dataset finalization (Figure~\ref{fig:longtail-pipeline}). Throughout curation, we retain the original issue and PR data and keep the reference patch unchanged.

During task construction, we identify \emph{FAIL\_TO\_PASS} (F2P) tests by applying the same evaluation test patch at the base commit and after the reference patch: these tests fail in the first state and pass in the second. \emph{PASS\_TO\_PASS} (P2P) tests pass in both states and serve as regression checks.

The solver receives the statement and repository environment and produces a solution patch. We evaluate each solution patch in a fresh, separate container, discarding solver changes to test files before applying the evaluation test patch. We use this evaluation procedure consistently across pipeline stages. An attempt is successful if its solution passes all designated F2P and P2P tests.

\subsection{Candidate Collection and Screening}
\label{sec:method-collection}

\paragraph{Source collection and prioritization.}
For each target language, we independently use the GitHub API to collect merged PRs linked to a single issue from eligible repositories. Candidates require a usable base commit and nonempty, separable reference and test patches. Appendix~\ref{sec:appendix-filtering} lists the repository eligibility criteria and PR filters. When processing capacity is limited, we prioritize larger reference patches by modified-file count, then patch length, both in descending order.

\paragraph{Semantic screening.}
\label{sec:method-screening}
Kimi K3~\cite{kimik3} assigns diagnostic categories inherited from SWE-rebench V2. We retain \texttt{A} (well-specified and solvable as-is) and two potentially repairable categories: \texttt{B2} (tests require names, signatures, or values absent from the statement) and \texttt{B5} (tests may depend on incidental reference-implementation details). Other categories, including tasks requiring unavailable external information, visual context, or non-English understanding, are excluded.

\subsection{Environment Construction and Initial Validation}
\label{sec:method-construction}

\paragraph{Environment preparation and initial audit.}
For each language, we prepare one or more versioned base Docker images to accommodate differences among repositories in runtime versions, build systems, and dependencies. Starting from the base commit, an auditor agent inspects the repository documentation, configuration, and CI, determines the project-native installation, build, and test commands, then installs task-specific dependencies and verifies the installation with existing tests. It also compares the problem statement, repository context, reference patch, and test patch, revising tests when necessary to generalize implementation-specific checks or add missing test coverage of the required behavior.

The auditor may replace implementation hints in the problem statement with \texttt{[HIDDEN]}, leaving other text unchanged. Examples of such replacements can be found in Appendix~\ref{sec:appendix-hidden-hints}. The subsequent repair procedure described in Section~\ref{sec:method-repair} is append-only. We preserve the raw issue, installation/build/test commands, and resulting test patch.

\paragraph{Reproducibility checks.}
We repeat installation, building, and testing from scratch in a clean container based on the task's selected base Docker image. We run the evaluation tests before and after applying the reference patch, requiring at least one F2P test and all designated evaluation tests to pass with the reference patch. We then build an image for each task with its repository dependencies pre-installed and repeat the checks inside it, comparing the results with the saved test groups and task configuration. Candidates enter the next stage only after these checks succeed; unsuccessful construction attempts are returned for revision or excluded.

\subsection{Solver-Guided Task Diagnosis and Repair}
\label{sec:method-repair}

\paragraph{Diagnosis from solution attempts.}
For each executable candidate, a solver attempts to complete the task. A judge then reviews the solver trajectories, solution patches, and evaluation outcomes in the task-specific Docker image, alongside the statement, reference patch, and tests. The verdict can expose underspecification or overly restrictive checks, while test-passing solutions that are still incorrect can reveal missing test coverage for edge cases.

The judge distinguishes implementation errors from description, test, or environment defects and accepts the candidate or requests repair. Once the judge accepts the current task version, the task proceeds to a judge agent using a different model configuration; if that judge identifies a new issue, the task returns to repair before judging continues. A task completes this process only after acceptance from all three judge models: GPT 5.6 Luna (xhigh), GPT 5.6 Terra (high), and Kimi K3 (high).

\paragraph{Permissible repairs.}
A separate repair agent receives the active judge's diagnosis and may revise the evaluation test patch and/or append clarifications to the problem statement. The statement produced by the initial task audit, including any \texttt{[HIDDEN]} replacements, remains immutable during repair. Clarifications are placed in an addendum (which may be revised during description repair) to specify the expected behavior while preserving the source issue's meaning and keeping the reference implementation undisclosed. Test changes aim to address false positive/negative cases, where wrong solutions might pass due to missed edge cases or incomplete checks, and correct solutions might fail due to overly restrictive or incorrect expectations. For description repair we use solver attempts from Kimi K3 (high), GPT 5.6 Terra (high), and GPT 5.6 Luna (xhigh); after the complete statement and its addendum are frozen for the test-repair phase, solver attempts use GPT 5.6 Sol (high). The full model configurations for the auditor, solver, judge, and repairer are provided in Appendix~\ref{app:prompts}, and the complete agent prompts are available in our code repository.

The environment remains fixed within this loop. Environment defects require returning to the construction stage. The repairer may not introduce production-code fixes through the test patch.

\paragraph{Validation and reuse of solution attempts.}
Revised tests must satisfy the execution criteria above at the base commit and with the unchanged reference patch. If a revision fails, we retain the last validated version and return error logs to the repairer. Tasks with faulty reference patches are discarded.

After validation, we save the updated task and return it to the solver--judge loop. Within this repair loop, test-only changes trigger reevaluation of existing solution patches, whereas changes to the problem statement require new solver attempts so that the solver can use the revised specification. The active judge reassesses the task using updated outcomes and available trajectories.

\paragraph{Iteration and audit history.}
We run diagnosis--repair--validation cycles in sets of up to five rounds until all three sequential judges accept the task or we identify a faulty reference patch and discard the task. Candidate processing for a language stops once at least 30 tasks are accepted for difficulty calibration. Across the 237 difficulty-selected tasks (see Section~\ref{sec:method-release} on difficulty calibration), repair takes 4.2 rounds on average (median 4, range 0--14). Figure~\ref{fig:repair-rounds-by-language} in Appendix~\ref{sec:appendix-repair-rounds} shows the full distribution by language. All intermediate revisions, trajectories, and verdicts are preserved.

\begin{table*}[t]
\centering
\caption{Overall performance and reported resource usage on Longtail-SWE. Success rates are percentages.}
\label{tab:overall-results}
\vskip 0.12in
\small
\setlength{\tabcolsep}{5pt}
\begin{tabular}{@{}lrrrrrr@{}}
\toprule
Model & \shortstack{Context\\limit} & \shortstack{Pass@1\\$\pm$ SEM} & Pass@3 & All-3 & \shortstack{Mean total\\tokens} & \shortstack{Mean\\steps} \\
\midrule
Opus 5.5 (high) & 1M & 81.52 $\pm$ 1.16 & 89.11 & 73.76 & 4.48M & 53.2 \\
GPT 6.1 Sol (high) & 1M & 78.38 $\pm$ 0.92 & 82.18 & 73.27 & 1.75M & 27.4 \\
GPT 5.6 Sol (high) & 1M & 75.74 $\pm$ 1.25 & 87.13 & 63.37 & 2.84M & 34.3 \\
GPT 5.6 Luna (xhigh) & 1M & 71.45 $\pm$ 0.44 & 84.16 & 56.93 & 13.01M & 76.5 \\
Kimi K3 (max) & 262K & 63.70 $\pm$ 2.43 & 76.24 & 48.51 & 16.84M & 151.4 \\
DeepSeek V4 Flash (high) & 1M & 56.11 $\pm$ 1.47 & 71.78 & 39.11 & 26.50M & 199.5 \\
GLM-5.3-Flash (max) & 262K & 48.18 $\pm$ 0.92 & 69.80 & 23.76 & 38.28M & 313.7 \\
Qwen 3.8 27B (xhigh) & 262K & 42.74 $\pm$ 0.92 & 62.38 & 20.79 & 23.06M & 214.4 \\
Qwen 3.6 27B (thinking) & 262K & 16.67 $\pm$ 0.83 & 23.76 & 9.41 & 7.20M & 113.4 \\
\bottomrule
\end{tabular}
\end{table*}

\subsection{Benchmark Finalization and Release}
\label{sec:method-release}

After completing repairs to task statements and tests, we calibrate task difficulty. For each judge-accepted task, we run Qwen 3.8 27B~\cite{qwen38} with xhigh reasoning three times using a 262K-token context window to guide difficulty-based selection. We exclude tasks solved in all three attempts unless doing so would leave fewer than 15 tasks in a language. In that case, we retain enough of these tasks to meet the minimum, prioritizing those with the longest average Qwen trajectory length across the three attempts. This minimum applies before post-evaluation validation; subsequent exclusions leave 12--27 tasks per language in the final 202-task benchmark.

\paragraph{Post-evaluation validation.}
\label{sec:method-post-evaluation-validation}
After evaluation, we refined the preliminary difficulty-selected pool of 237 tasks through an iterative audit, repair, and exclusion process informed by 5,688 stored solver attempts from eight model configurations, with three attempts per configuration per task (see Section~\ref{sec:experiments-evaluation-protocol}). We first corrected evaluator defects in patch application and test-output parsing, then replayed the stored solution patches. We subsequently examined problem statements, tests, reference implementations, solver patches, and execution logs to identify tests that rejected valid solutions or accepted solutions that violated stated requirements. This semi-automated process combined human review with assistance from GPT 6 Luna, Claude Opus 5.5, GPT 6 Astra, and GPT 6 Sol, which also helped refine the tests and problem descriptions. We checked test changes by replaying stored solution patches and ran new solver attempts when problem statements changed. We excluded tasks that could not be repaired without changing their intent, such as those whose tests left requested functionality untested or conflicted with the reference patch.
We retained only tasks that passed revalidation, excluding 35 tasks and yielding the final benchmark of 202 tasks. Table~\ref{tab:funnel} (Section~\ref{sec:ablations-funnel}) presents the full selection funnel.

\paragraph{Data and code availability.}
The complete release is publicly available through \url{https://github.com/SWE-rebench/Longtail-SWE}, with all materials provided either directly in the repository or through links to external resources. This includes the 202-task Longtail-SWE benchmark, the easier companion set, prompts, reproduction scripts, all benchmark instances, original and curated problem descriptions and tests, source metadata and reference patches, environment configurations, the complete data-collection, environment-construction, audit, repair, and evaluation tooling, all agent execution trajectories, and benchmark container images.
Our original code is licensed under the MIT License, and our original dataset contributions under CC BY 4.0; third-party materials remain subject to their original licenses.

\begin{table*}[t]
\centering
\caption{Pass@1 (\%) by programming language, averaged over three attempts per task.}
\label{tab:language-results-main}
\vskip 0.12in
\small
\setlength{\tabcolsep}{3pt}
\resizebox{\textwidth}{!}{
\begin{tabular}{@{}lrrrrrrrrrr@{}}
\toprule
Language & $N$
& \shortstack{Opus 5.5\\(high)} 
& \shortstack{GPT 6.1\\Sol (high)}
& \shortstack{GPT 5.6\\Sol (high)}
& \shortstack{GPT 5.6\\Luna (xhigh)}
& \shortstack{Kimi K3\\(max)}
& \shortstack{DeepSeek V4\\Flash (high)}
& \shortstack{GLM-5.3-Flash\\(max)}
& \shortstack{Qwen 3.8 27B\\(xhigh)}
& \shortstack{Qwen 3.6 27B\\(thinking)} \\
\midrule
Clojure & 13 & \textbf{82.1} & 74.4 & 51.3 & 46.2 & 64.1 & 35.9 & 30.8 & 38.5 & 10.3 \\
Dart    & 14 & 71.4 & 61.9 & \textbf{81.0} & \textbf{81.0} & 66.7 & 61.9 & 40.5 & 38.1 & 11.9 \\
Elixir  & 23 & 75.4 & \textbf{81.2} & 72.5 & 65.2 & 53.6 & 34.8 & 31.9 & 30.4 & 8.7 \\
Erlang  & 12 & \textbf{100.0} & \textbf{100.0} & 86.1 & 83.3 & 83.3 & 75.0 & 50.0 & 58.3 & 25.0 \\
F\#     & 16 & 66.7 & 62.5 & \textbf{81.3} & 68.8 & 47.9 & 41.7 & 43.8 & 29.2 & 10.4 \\
Haskell & 14 & 92.9 & \textbf{95.2} & 83.3 & 83.3 & 76.2 & 78.6 & 66.7 & 52.4 & 26.2 \\
Julia   & 27 & 86.4 & \textbf{93.8} & 85.2 & 77.8 & 76.5 & 58.0 & 59.3 & 38.3 & 16.0 \\
Kotlin  & 12 & \textbf{77.8} & 63.9 & 72.2 & 61.1 & 50.0 & 66.7 & 58.3 & 47.2 & 22.2 \\
Lua     & 13 & \textbf{89.7} & \textbf{89.7} & \textbf{89.7} & 74.4 & 66.7 & 66.7 & 59.0 & 51.3 & 17.9 \\
OCaml   & 13 & \textbf{84.6} & 76.9 & 64.1 & 76.9 & 56.4 & 53.8 & 43.6 & 28.2 & 5.1 \\
R       & 14 & 78.6 & 78.6 & 76.2 & 69.0 & \textbf{81.0} & 69.0 & 57.1 & 69.0 & 42.9 \\
Scala   & 18 & \textbf{87.0} & 66.7 & 66.7 & 70.4 & 53.7 & 50.0 & 37.0 & 44.4 & 14.8 \\
Swift   & 13 & \textbf{69.2} & 64.1 & \textbf{69.2} & \textbf{69.2} & 51.3 & 56.4 & 53.8 & 46.2 & 12.8 \\
\midrule
Overall & 202 & \textbf{81.5} & 78.4 & 75.7 & 71.5 & 63.7 & 56.1 & 48.2 & 42.7 & 16.7 \\
\bottomrule
\end{tabular}
}
\end{table*}

\section{Experiments}
\label{sec:experiments}

\subsection{Experimental Setup}

\paragraph{Models and evaluation protocol.}
\label{sec:experiments-evaluation-protocol}

We evaluate nine model configurations on the 202-task Longtail-SWE: Opus 5.5 (high), GPT 6.1 Sol (high), GPT 5.6 Sol (high), GPT 5.6 Luna (xhigh), Kimi K3 (max), DeepSeek V4 Flash (high), GLM-5.3-Flash (max), Qwen 3.8 27B (xhigh), and Qwen 3.6 27B (thinking). Each configuration uses a SWE-agent-like scaffold for three attempts per task. We compare these specific model--agent configurations, with reasoning modes and context limits listed in Table~\ref{tab:overall-results}. The agent prompts are available in the repository.

\paragraph{Metrics.}
An attempt succeeds when its solution passes all required F2P and P2P tests under the protocol in Section~\ref{sec:method-overview}. Pass@1 averages success across tasks and attempts; Pass@3 is the fraction of tasks solved at least once, and All-3 the fraction solved in all three attempts. We report success rates as percentages, run-level standard error of the mean (SEM) for Pass@1, and mean total token usage and agent step counts.

\paragraph{Selection and evaluation.}
As described in Section~\ref{sec:method-release}, difficulty selection uses three Qwen 3.8 27B calibration attempts and excludes tasks solved in all three unless needed to retain at least 15 tasks per language before post-evaluation validation. The Qwen 3.8 results in Table~\ref{tab:overall-results} use a different set of three attempts; no attempt used for difficulty filtering is included in these reported scores. Nevertheless, because task selection depended partly on calibration runs of the same model, the reported Qwen results may still be biased by model-dependent selection and the language-coverage constraint.
\subsection{Dataset Composition and Task Scope}
\label{sec:dataset-composition}

Longtail-SWE contains 202 tasks from 129 repositories across 13 programming languages. Core feature development is the most frequent task-type annotation (67.3\%), followed by major bug fixes (27.2\%) and integration features; labels are not mutually exclusive. Appendix~\ref{sec:appendix-composition} provides the full task-type distribution in Table~\ref{tab:task-categories}, language and repository counts, and LLM-annotated difficulty.
Appendix~\ref{sec:appendix-benchmark-comparison} compares coverage and structural statistics with other repository-level benchmarks.
\subsection{Overall Results}

\paragraph{Task resolution.} Opus 5.5 (high) has the highest point estimates on all three success metrics, with 81.52\% Pass@1, exceeding runner-up GPT 6.1 Sol by 3.14 percentage points (Table~\ref{tab:overall-results}). Even after three attempts, 10.89\% of tasks remain unsolved by Opus. Qwen 3.8 27B exceeds Qwen 3.6 27B on all three success metrics.

\paragraph{Resource usage.} GPT 6.1 Sol has the lowest reported mean token usage and step count, followed by GPT 5.6 Sol and Opus. Luna uses approximately three times as many tokens as Opus. Kimi uses more tokens and steps than Luna despite a lower Pass@1, while GLM-5.3-Flash has the highest means for both resources.

\begin{figure}[tb]
  \centering
  \includegraphics[width=0.6\linewidth]{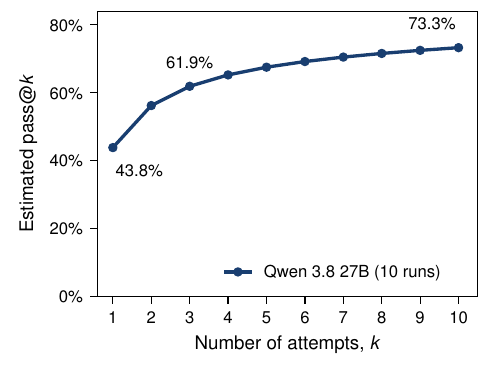}
  \caption{Qwen 3.8 27B pass@$k$ on the 202-task subset, estimated from ten runs per task. Pass@1 is 43.8\% and pass@10 is 73.3\%.}
  \label{fig:qwen-pass-at-k}
\end{figure}

\paragraph{Repeated attempts.}
Figure~\ref{fig:qwen-pass-at-k} reports Qwen 3.8 27B Pass@k estimates on the final 202-task benchmark using ten attempts per task. Pass@k increases from 43.8\% at $k=1$ to 73.3\% at $k=10$, with diminishing marginal gains as $k$ increases. Many tasks are therefore solvable by Qwen but not reliably in a single attempt, while 26.7\% remain unsolved after ten attempts.

\subsection{Performance Across Languages}
\label{sec:language-failure-analysis}

\paragraph{Variation across language subsets.}
Table~\ref{tab:language-results-main} reports per-language results for all nine configurations. Averaging Pass@1 equally across configurations, Erlang has the highest mean success rate (73.5\%), followed by Haskell (72.8\%) and R (69.0\%); Clojure, F\#, and Elixir have the lowest rates (48.1\%, 50.2\%, and 50.4\%). The highest and lowest means differ by approximately 1.5-fold.
Opus 5.5 leads outright in four language subsets, ties GPT 6.1 Sol on Erlang, ties both Sol configurations on Lua, and ties GPT 5.6 Sol and Luna on Swift. GPT 6.1 Sol leads on Elixir, Haskell, and Julia. GPT 5.6 Sol leads on F\# and ties Luna on Dart, and Kimi K3 leads on R (Table~\ref{tab:language-results-main}).

\subsection{Human Validation of the Judge}
\label{sec:human-validation}

\paragraph{Audit set and annotation protocol.}
We evaluate the judge on an audit set of 65 tasks stratified by language and pipeline stage, used exclusively for evaluation rather than judge development. Two engineers independently assessed each task against predefined criteria using the task artifacts, solver trajectory, and evaluation test patch, with optional agent assistance for code inspection. Every decision required a written justification, including concrete evidence whenever tests rejected valid solutions or accepted incorrect ones.

The two annotators marked 44 and 41 of the 65 tasks as problematic, respectively, agreeing on 40 tasks (Cohen's $\kappa = 0.15$, reflecting the difficulty of identifying subtle task defects). They then reconciled their notes to establish a consensus gold label for every task. Of the 25 initial disagreements, 21 were resolved as \texttt{REJECT}, typically because one annotator spotted a defect missed by the other. The remaining four received \texttt{ACCEPT}: one flagged test omission involved behavior outside the problem statement, one concerned a negligible corner case, and two involved non-standard test-result parsing that proved valid. The final gold set comprises 14 accepted and 51 rejected tasks.

\paragraph{Judge setup and results.}
After freezing the gold labels, we ran our sequential three-judge protocol (Section~\ref{sec:method-repair}) using GPT 5.6 Luna (xhigh), GPT 5.6 Terra (high), and Kimi K3 (high), which returns \texttt{ACCEPT} only upon unanimous acceptance and routes any \texttt{REJECT} to repair to prioritize sensitivity to defects. The protocol rejected 56 tasks in total, correctly identifying 49 of the 51 problematic gold tasks. Treating rejection as the positive class, the judge achieves $87.5\%$ precision and $96.1\%$ recall: only $3.9\%$ of human-identified defective tasks were missed, while $12.5\%$ of rejected tasks were sent to repair unnecessarily.

\begin{table*}[b]
\centering
\caption{Ablation of semantic-screening scope and the construction agent. Strict EV rates use the screened candidates in each row as the denominator.}
\label{tab:audit-agent-ablation}
\vskip 0.12in
\small
\begin{tabular}{@{}lrrr@{}}
\toprule
Pipeline configuration & Candidates & Strict EV passes & Pass rate (\%) \\
\midrule
SWE-rebench V2 install agent, \texttt{A} only & 1,096 & 377 & 34.4 \\
Combined auditor, \texttt{A} only & 1,096 & 539 & 49.2 \\
Combined auditor, \texttt{A}/\texttt{B2}/\texttt{B5} & 2,640 & 1,247 & 47.2 \\
\bottomrule
\end{tabular}
\end{table*}

\section{Ablations}
\label{sec:ablations}

\subsection{Construction-Pipeline Ablation}
\label{sec:experiments-funnel}

Long-tail language ecosystems offer fewer eligible pull requests, so early construction losses can leave too few tasks for meaningful per-language evaluation. We therefore compare retaining potentially repairable candidates and combining environment installation with test auditing against the SWE-rebench V2 pipeline, then track retention through iterative repair and difficulty selection.

\paragraph{Retaining repairable annotations.}

SWE-rebench V2 retains only candidates labeled \texttt{A}, whose statements and tests appear aligned. We also retain potentially repairable \texttt{B2} candidates (tests require names, signatures, or values absent from the statement) and \texttt{B5} candidates (tests may enforce incidental implementation details), expanding the pool from 1,096 to 2,640 tasks. Appendix~\ref{sec:appendix-construction-details} gives the category counts.

\paragraph{Combined environment and test audit.}
Our Kimi K3 auditor~\cite{kimik3} combines environment installation with test-patch repair (Section~\ref{sec:method-construction}). Strict execution validation (EV) requires the tests to compile and execute in both repository states, with at least one F2P transition and no regressions after the reference patch.

Table~\ref{tab:audit-agent-ablation} compares construction configurations on the same 1,096 \texttt{A} candidates: strict EV passes increase from 377 to 539 (34.4\% to 49.2\%). The auditor modifies test patches in 331 of the 539 successful \texttt{A} tasks (61.4\%). These construction-stage revisions go beyond environment installation and precede iterative solver--judge--repair. Expanding the input pool to include \texttt{B2}/\texttt{B5} candidates increases the yield from 539 to 1,247 passes, adding 708 passes from outside the original \texttt{A} pool.

The combined screening and construction changes increase strict EV passes from 6 to 86 for Kotlin, 2 to 78 for Scala, and 9 to 54 for Swift. Appendix~\ref{sec:appendix-construction-details} provides the remaining per-language results (Table~\ref{tab:ev-passes-by-language}) and discusses the recovered tasks.

\paragraph{End-to-end funnel.}
\label{sec:ablations-funnel}

Table~\ref{tab:funnel} tracks both paths from a common candidate pool. Iterative repair and quality review yield 533 accepted tasks; difficulty selection retains 237 tasks, and the remaining tasks form the easier companion set. Post-evaluation validation reduces the 237 difficulty-selected tasks to the final 202-task Longtail-SWE benchmark (Section~\ref{sec:method-post-evaluation-validation}). Only 68 of the benchmark tasks would remain available through the original funnel.

\definecolor{comparisongray}{gray}{0.45}
\definecolor{comparisonbg}{gray}{0.95}

\begin{table*}[t]
\caption{Construction funnel from a common pool of 3,261 candidates. Parentheses report the percentage of the initial pool retained at each stage. Gray entries are shown only for downstream comparison and are not part of the SWE-rebench V2 construction pipeline.}
\label{tab:funnel}
\vskip 0.12in
\centering
\begin{tabular}{lrr}
\toprule
Stage & SWE-rebench V2 path & Longtail-SWE \\
\midrule
Before semantic annotation & 3,261 (100.0\%) & 3,261 (100.0\%) \\
After semantic screening & 1,096 (33.6\%) & 2,640 (81.0\%) \\
Strict execution validation & 377 (11.6\%) & 1,247 (38.2\%) \\
Image build and revalidation & 364 (11.2\%) & 1,201 (36.8\%) \\
After repair quality review & \cellcolor{comparisonbg}\textcolor{comparisongray}{197 (6.0\%)} & 533 (16.3\%) \\
Difficulty-selected pool & \cellcolor{comparisonbg}\textcolor{comparisongray}{80 (2.5\%)} & 237 (7.3\%) \\
After post-evaluation validation & \cellcolor{comparisonbg}\textcolor{comparisongray}{68 (2.1\%)} & 202 (6.2\%) \\
Final re-annotation as A & \cellcolor{comparisonbg}\textcolor{comparisongray}{66 (2.0\%)} & 196 (6.0\%) \\
\bottomrule
\end{tabular}
\end{table*}

\paragraph{Post-repair annotation.}

After repair and post-evaluation validation, 196 of the 202 tasks (97.0\%) receive an \texttt{A} annotation, compared with 66 of the 68 tasks available through the SWE-rebench V2 path. Counting only final \texttt{A} annotations on both paths still yields 2.97 times as many benchmark tasks (196 vs.\ 66; Table~\ref{tab:funnel}). Appendix~\ref{sec:appendix-construction-details} reports results for the full accepted pool.

\paragraph{Judge-based quality audit of the SWE-rebench V2 yield.}

Additionally, we evaluated all 364 tasks that passed image build and revalidation through the SWE-rebench V2 construction path using a single GPT 5.6 Luna (xhigh) judge. The judge identified at least one task-quality issue in 328 tasks (90.1\%). The issue categories are not mutually exclusive: 300 tasks (82.4\%) had hidden tests that required undocumented APIs or exact outputs, and 323 (88.7\%) had insufficient test coverage. Only 36 tasks (9.9\%) were accepted by this single judge. Because our sequential protocol accepts a task only when all three judges accept it, applying the remaining judges could only reduce this accepted set; 36 is therefore an upper bound on the number of these tasks that would pass the full protocol without repair. In contrast, every task released in Longtail-SWE was accepted by all three judges after the repair process.

\subsection{Pre- versus Post-Repair Pass Rates}

For 63 released tasks, the problem description did not change during repair. We therefore re-evaluated the solution patches from the same three Opus 5.5 (high) trajectories per task against both the pre-repair and post-repair tests. Table~\ref{tab:pre-post-repair-pass-rates} shows that Pass@1 increases by 38.1 percentage points after repair, while Pass@3 increases by 41.3 points and All-3 by 33.3 points. Because the trajectories and problem descriptions are held fixed, these differences reflect changes to the evaluation tests. Appendix~\ref{sec:appendix-test-repair-case-studies} presents two case studies from this subset; Appendix~\ref{sec:appendix-statement-repair-case-study} separately illustrates a description-only repair with unchanged tests.

\begin{table}[t]
\centering
\caption{Opus 5.5 pre- and post-repair performance on the 63 tasks with unchanged problem descriptions.}
\label{tab:pre-post-repair-pass-rates}
\vskip 0.12in
\small
\setlength{\tabcolsep}{3pt}
\begin{tabular}{@{}lrrr@{}}
\toprule
Evaluation tests & Pass@1 $\pm$ SEM & Pass@3 & All-3 \\
\midrule
Pre-repair & $29.1 \pm 2.8$ & 36.5 & 20.6 \\
Post-repair & $67.2 \pm 3.5$ & 77.8 & 54.0 \\
\bottomrule
\end{tabular}
\end{table}

The paired comparison also holds across ten Qwen 3.8 27B attempts on the same 63 tasks: Pass@1 rises from 15.2\% to 31.0\%, and Pass@10 rises from 36.5\% to 68.3\%.
\section{Discussion}
\label{sec:discussion}

Longtail-SWE broadens repository-level evaluation across 13 language ecosystems that are underrepresented in existing benchmarks. Our results provide a starting point for studying coding agents in these ecosystems.

Revising task descriptions and tests helps recover tasks from a limited candidate pool. Our fixed-patch ablation isolates the effect of test refinement on a subset with unchanged descriptions. Further ablations are needed to separate the effects of other refinement steps and assess whether construction-time solver choices favor particular models.

The human audit evaluates judge reliability on a small sample drawn from multiple pipeline stages. It does not establish the error rate of the final benchmark; this requires an independent audit of the released tasks.

\subsection{Threats to Validity}
\label{sec:threats}

\paragraph{Construct validity.}
Fixing the reference patch and allowing only appended clarifications limit changes to task intent, but revisions may still alter the task contract. The paired analysis of 63 tasks with unchanged descriptions supports the test revisions; description changes are not separately evaluated in that analysis. The 65-task human audit shows high recall for problematic tasks (49/51, 96.1\%) but low specificity for acceptable tasks (7/14, 50.0\%), alongside low initial annotator agreement. This low specificity sends some acceptable tasks through unnecessary repair, increasing the computational resources required by the construction process.

\paragraph{Internal validity.}
Difficulty selection with Qwen 3.8 27B may bias its reported scores. Some evaluated models also participate in construction and may benefit from the resulting task interpretations, though models not used during construction offer a partial check. In future work, we plan to investigate how a model’s involvement in benchmark construction affects its scores.

\paragraph{Contamination.}
Tasks derive from public issues and pull requests, so overlap with training data cannot be ruled out. Because training cutoffs are often undisclosed and not determined by release dates, it is difficult to ensure that a fixed task set postdates the training data of all evaluated models. However, our pipeline can collect tasks created on or after a specified date, restricting evaluation to tasks created after a given model's training cutoff.

\paragraph{External validity.}
Repository and PR filters (permissive licenses, minimum popularity, single-issue links, prioritization of larger patches) shape the sample. With 12--27 tasks per language, language-level differences reflect these tasks and the evaluated configurations rather than inherent language difficulty.

\section{Conclusion}
\label{sec:conclusion}

We introduced Longtail-SWE, a benchmark of 202 issue-resolution tasks from 129 repositories across 13 underrepresented programming languages. Building on SWE-rebench V2, we developed a solver--judge--repairer pipeline designed to retain challenging candidates by addressing defects in task descriptions and tests. Across nine evaluated configurations, the highest observed scores are 81.52\% Pass@1 and 89.11\% Pass@3. The benchmark supports evaluation across these ecosystems and further research on reliable benchmark construction; current data and code availability is described in Section~\ref{sec:method-release}.

\bibliographystyle{nebius}
\bibliography{sections/references}

\clearpage
\appendix

\renewcommand{\thefigure}{A\arabic{figure}}
\renewcommand{\thetable}{A\arabic{table}}
\renewcommand{\thelstlisting}{A\arabic{lstlisting}}
\lstdefinestyle{problemstatement}{
  basicstyle=\ttfamily\footnotesize,
  breaklines=true,
  breakatwhitespace=true,
  columns=fullflexible,
  keepspaces=true,
  showstringspaces=false,
  frame=single,
  float=*t 
}

\section{Appendix}
\label{sec:appendix}

\subsection{Detailed Collection and Filtering Rules}
\label{sec:appendix-filtering}

\paragraph{Repository-level collection and filtering rules.}
Repository collection uses the following criteria and search procedures:
\begin{itemize}
    \item Repositories are searched separately for each target language using the GitHub language label.
    \item Repositories must have at least 10 GitHub stars.
    \item Repositories must have been updated on or after January 1, 2026.
    \item Archived repositories are excluded.
    \item Disabled repositories are excluded.
    \item Only licenses on the permissive SPDX allowlist are accepted: 0BSD, Apache-2.0, BSD-2-Clause, BSD-3-Clause, BSL-1.0, CC0-1.0, ISC, MIT, PostgreSQL, Python-2.0, Unlicense, and Zlib.
    \item Missing or unknown licenses are rejected unless an exception is explicitly recorded.
    \item No filter is specified in the default configuration for total file count, lines of code, or checkout size.
    \item Searches reaching the 1,000-result cap are split into nonoverlapping date or star-count ranges; results are then deduplicated.
\end{itemize}

\paragraph{Pull-request-level collection and filtering rules.}
PR collection uses the following criteria, with processing order noted separately from eligibility:
\begin{itemize}
    \item Only merged PRs are accepted.
    \item Each PR must be linked to an issue.
    \item PRs linked to multiple distinct issues are excluded; duplicate records of the same PR--issue pair do not count as multiple issues.
    \item PRs must fall within the configured inclusive date window.
    \item Bot-generated PRs are excluded.
    \item Automated dependency updates are excluded.
    \item PRs may change at most 1,000 files.
    \item The production/reference patch must modify at least one supported file.
    \item The test patch must modify at least one supported source or test-bearing file.
    \item The production patch must contain at least five added lines by default.
    \item The combined size of recognized diff sections for code files must be less than 100,000 characters.
    \item The problem statement must be less than 100,000 characters.
    \item No upper limit on test patch size is imposed by the default language-specific procedure.
    \item Both the reference patch and the test patch must be nonempty.
    \item A usable base commit must exist.
    \item Candidates are rejected if tests are detected inside production (non-test) files.
    \item Some language batches are sorted by change size; this determines processing order rather than serving as a universal acceptance filter.
\end{itemize}

\subsection{Additional Construction-Pipeline Results}
\label{sec:appendix-construction-details}

\paragraph{Semantic-screening counts.}
The common pre-screening pool contains 3,261 candidates across 13 languages. Of these, 1,096 receive label \texttt{A}, 1,408 receive \texttt{B2}, and 136 receive \texttt{B5}, as defined in Section~\ref{sec:experiments-funnel}. Retaining all three categories adds 1,544 candidates to the \texttt{A}-only pool, a 140.9\% increase.

\paragraph{Test-patch revisions during construction.}
Among the 539 \texttt{A} candidates that pass strict execution validation with the combined auditor, 331 (61.4\%) have test patches modified by the auditor. These revisions occur during construction, before iterative solver--judge--repair.

\paragraph{Execution-validation passes by language.}
Table~\ref{tab:ev-passes-by-language} breaks down strict EV passes by language for the construction configurations in Section~\ref{sec:experiments-funnel}. Relative increases are particularly large for Kotlin, Scala, and Swift. Audit logs indicate that repairing test patches that failed to compile at the base revision recovered tasks rejected by the installation-only baseline.

\begin{table}[t]
\centering
\caption{Execution-validation (EV) passes by language. \mbox{SWE-rebench V2} uses the \texttt{A}-only pool; Longtail-SWE uses the expanded \texttt{A}/\texttt{B2}/\texttt{B5} pool and combined auditor. The comparison therefore includes both screening and construction changes.}
\label{tab:ev-passes-by-language}
\vskip 0.12in
\small
\begin{tabular}{@{}lrr@{}}
\toprule
Language & SWE-rebench V2 & Longtail-SWE \\
\midrule
Clojure & 21 & 55 \\
Dart    & 20 & 122 \\
Elixir  & 59 & 109 \\
Erlang  & 36 & 57 \\
F\#     & 62 & 74 \\
Haskell & 49 & 126 \\
Julia   & 27 & 123 \\
Kotlin  & 6  & 86 \\
Lua     & 42 & 153 \\
OCaml   & 16 & 114 \\
R       & 28 & 96 \\
Scala   & 2  & 78 \\
Swift   & 9  & 54 \\
\midrule
\textbf{Total} & \textbf{377} & \textbf{1,247} \\
\bottomrule
\end{tabular}
\end{table}

\begin{table*}[b]
\centering
\caption{Coverage and structural statistics for selected repository-level benchmarks. Tasks, repositories, and languages are counts; all remaining columns are per-task means. Words denotes statement words, files denotes changed solution files, and reference/test additions count added lines in the respective patches. Dashes indicate unavailable measurements.}
\label{tab:benchmark-comparison}
\vskip 0.12in
\small
\setlength{\tabcolsep}{3pt}
\begin{tabular}{@{}lrrrrrrrrr@{}}
\toprule
Dataset & Tasks & Repos. & Langs. & Words & Files & \shortstack{Reference\\additions} & \shortstack{Test\\additions} & F2P & P2P \\
\midrule
Longtail-SWE & 202 & 129 & 13 & 513.25 & 13.14 & 414.82 & 645.87 & 23.97 & 42.91 \\
SWE-Bench ProMax & 170 & 70 & 7 & 481.45 & 12.38 & 342.05 & 236.52 & --- & --- \\
SWE-Bench Pro (Public) & 731 & 11 & 4 & 178.88 & 5.07 & 120.34 & 105.73 & 14.43 & 57.46 \\
SWE-bench Multilingual & 300 & 41 & 9 & 234.83 & 1.74 & 28.87 & 39.35 & 1.44 & 34.02 \\
\bottomrule
\end{tabular}
\end{table*}

\paragraph{Post-repair annotation of the accepted pool.}
Of the 533 tasks accepted after iterative repair and quality review, 517 (97.0\%) receive a final \texttt{A} annotation; annotation cannot be completed for five additional tasks because they exceed the annotator's context limit. Difficulty selection retains a preliminary pool of 237 tasks; the remaining tasks form the easier companion set. Post-evaluation validation then reduces this pool to the final 202-task benchmark (Section \ref{sec:method-post-evaluation-validation}).

\subsection{Structural Comparison with Other Benchmarks}
\label{sec:appendix-benchmark-comparison}

Among the benchmarks in Table~\ref{tab:benchmark-comparison}, Longtail-SWE covers the most repositories and languages and has larger reference changes and test additions on average.

\subsection{Implementation Hints Hidden by the Auditor Agent}
\label{sec:appendix-hidden-hints}

In the final 202-task dataset, the initial task audit replaced parts of the problem description with \texttt{[HIDDEN]} in 56 tasks (113 occurrences in total).

We sampled two examples of such replacements, which are given in Listings~\ref{lst:hidden-hint-1}~and~\ref{lst:hidden-hint-2}.

\begin{lstlisting}[style=problemstatement,caption={Hidden implementation hint in livekit\_\_client-sdk-android-874; the replaced content is given after three dashes.},label={lst:hidden-hint-1}]
Library plants Timber.DebugTree() globally, affecting host app logging
Hi!

LiveKit currently plants Timber.DebugTree() internally when loggingLevel != OFF
(if no DebugTree is present in the forest).

Since Timber is a global logger, this affects the entire host application,
not just LiveKit logs, and removes control over logging configuration
(e.g. custom Trees like Crashlytics-only logging).

Library code generally should not plant Timber trees, as this is an
application-level concern.

[HIDDEN]

Thanks!
---
Suggested alternatives:
- Do not plant any Timber Tree, only log via Timber with a specific tag
- Provide a logger interface / callback so the host app decides where logs go
- Expose a flag or API to opt out of planting DebugTree
\end{lstlisting}

\begin{lstlisting}[style=problemstatement,caption={Hidden implementation hint in amazon-mq\_\_rabbitmq-stream-s3-205; the replaced content is given after three dashes.},label={lst:hidden-hint-2}]
AWS API: Use a `gen_server` for short-lived credentials
Currently `rabbitmq_stream_s3_api_aws` uses a lock (`global:set_lock/3`) to prevent multiple processes from attempting to refresh short-lived credentials simultaneously. We should replace this with a simple `gen_server` which is responsible for holding and refreshing short-lived credentials. Using a `gen_server` would avoid the locking (which is unnecessarily complicated) and it would let us eagerly refresh the short-lived credentials in the background when the credentials are close to expiration. [HIDDEN]
---
It can start up and return `ignore` from `c:gen_server:init/1` when long-lived credentials are configured.
\end{lstlisting}

\subsection{Case Studies of Test Repair}
\label{sec:appendix-test-repair-case-studies}

We sampled two illustrative cases from the 63 released tasks whose problem statements did not change during repair. This restriction allows the same solver trajectories and solution patches to be evaluated against the pre-repair and post-repair tests: for each corresponding run, the Opus 5.5 (high) trajectory and submitted patch are identical across the two evaluations. The cases illustrate two distinct test defects and are not intended to estimate their prevalence in the dataset.

\paragraph{Removing an unstated representation constraint.}
The \texttt{esl\_\_amoc-193} task asks that a required module attribute accept a binary description, making Elixir strings usable alongside the existing Erlang charlist form. All three Opus patches add a binary clause to \texttt{check\_mandatory\_fields} while retaining the existing charlist clause. The pre-repair tests correctly check that a binary description is accepted, but also require a charlist description to be stored as a binary and make additional exact-value assertions about nested character data and Unicode. The problem statement does not require conversion of existing charlists. All three patches fail the pre-repair suite at the charlist assertion even though they implement the requested binary-acceptance behavior.

The post-repair tests continue to require the binary description to be preserved exactly. For the legacy charlist, however, they check successful processing without prescribing its stored representation; they also remove the nested-character-data and Unicode-normalization assertions. The test identifiers and command remain unchanged. Under the post-repair tests, the same three Opus patches pass. This repair prevents rejection of solutions that meet the stated requirement but do not reproduce the reference patch's normalization choice.

\paragraph{Adding coverage for an incomplete implementation.}
The \texttt{intersectmbo\_\_cardano-ledger-5671} task asks for \texttt{NoThunks} instances to be removed from predicate failures throughout the codebase and from \texttt{ContextError} types in every era. The run~2 Opus patch removes many such instances but leaves five in the TPraos protocol code. The reference patch removes all five.

The pre-repair tests check only five Shelley predicate-failure types and a \texttt{Void} control case, so they do not exercise the remaining TPraos types and incorrectly accept the run~2 patch. The post-repair tests expand the checks across eras and protocol rules while retaining a control for the detection mechanism. When the identical run~2 patch is evaluated against the post-repair suite, 5 of 74 examples fail---exactly the five remaining TPraos instances. The test command is unchanged. Thus, the post-repair tests turn a false positive into a failure by covering behavior explicitly required by the problem statement.

\begin{figure*}[t]
    \centering
    
    \includegraphics[width=\linewidth,height=0.45\textheight,keepaspectratio]{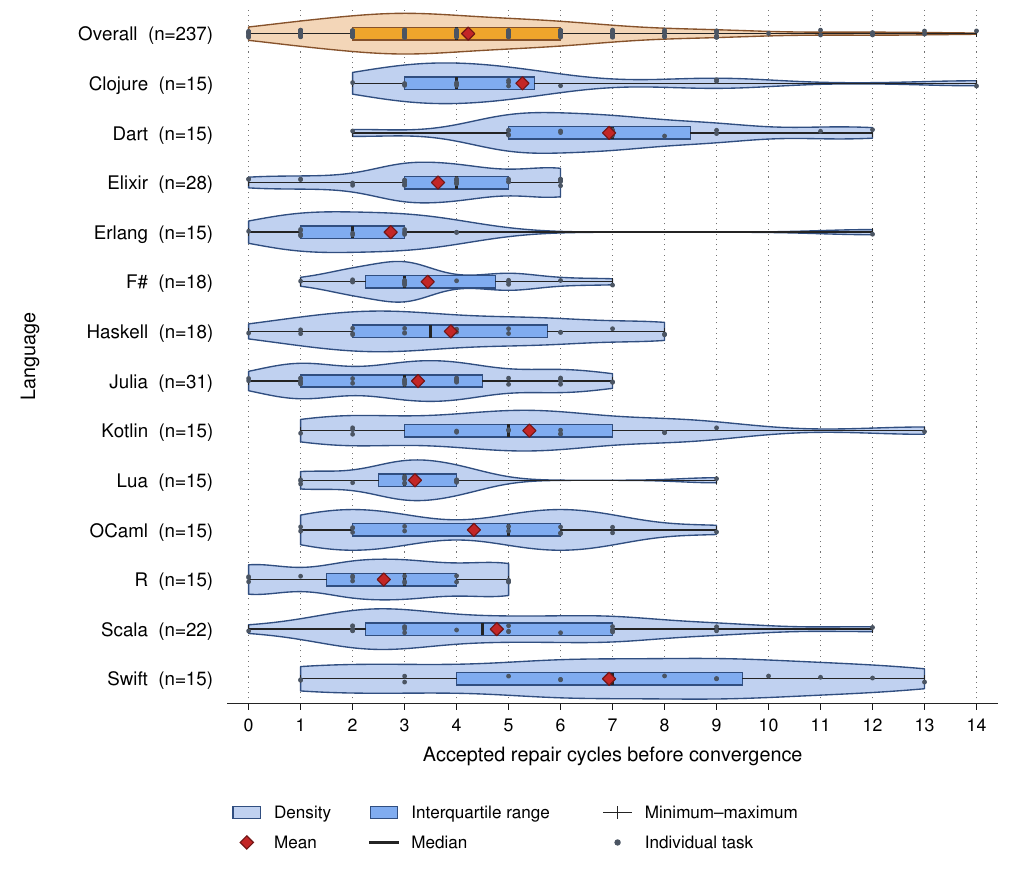}
    \caption{Distribution of accepted repair rounds before convergence for the 237 difficulty-selected tasks, overall and by language (overall mean: 4.22 rounds).}
    \label{fig:repair-rounds-by-language}
\end{figure*}

\subsection{Case Study of Problem-Statement Repair}
\label{sec:appendix-statement-repair-case-study}

We additionally sampled one of the five tasks for which repair changed only the problem statement, leaving the evaluation \texttt{test\_patch} byte-for-byte unchanged. This complements the preceding test-repair cases by illustrating how statement repair can expose requirements that the tests already enforce. The selected task, \texttt{penelopeysm\_\_flexichains.jl-229}, asks for merging \texttt{FlexiSummary} objects.
Its original 68-word statement describes the central use case, shown in Listing~\ref{lst:flexisummary-before}. The repaired 401-word statement in Listing~\ref{lst:flexisummary-after} retains that text and makes the tested contract explicit.

The unchanged tests cover rightmost-value precedence, missing entries for absent key--statistic combinations, retained- and collapsed-axis behavior, mixed key types and their warning, and merges of three or more summaries. The original statement communicates the main operation, whereas the expanded statement surfaces edge cases and design choices that a solver could not reliably infer.

\subsection{Repair-Round Distribution}
\label{sec:appendix-repair-rounds}

Figure~\ref{fig:repair-rounds-by-language} shows the distribution of repair rounds overall and by language.

\subsection{Detailed Dataset Composition}
\label{sec:appendix-composition}

Table~\ref{tab:task-categories} reports the full distribution of task-type annotations. Labels are not mutually exclusive: a task contributes to every category assigned to it.

\begin{table}[t]
\centering

\caption{Task-type annotations for the 202 Longtail-SWE tasks. A task may have multiple labels; percentages use all 202 tasks as the denominator and do not sum to 100\%.}
\label{tab:task-categories}
\vskip 0.12in
\small
\begin{tabular}{@{}lrr@{}}
\toprule
Task type & Tasks & Share (\%) \\
\midrule
Core features & 136 & 67.3 \\
Major bugs & 55 & 27.2 \\
Integration features & 25 & 12.4 \\
Edge-case bugs & 20 & 9.9 \\
Minor bugs & 19 & 9.4 \\
UI/UX features & 11 & 5.4 \\
DevOps enhancements & 9 & 4.5 \\
Performance bugs & 8 & 4.0 \\
Security bugs & 5 & 2.5 \\
Critical bugs & 4 & 2.0 \\
Documentation enhancements & 4 & 2.0 \\
Regression bugs & 1 & 0.5 \\

\bottomrule
\end{tabular}
\end{table}
Table~\ref{tab:language-distribution} breaks down the final sample by language and repository count. Difficulty labels are LLM annotations rather than measurements of human resolution time.

\begin{table}[t]
\centering
\caption{Distribution of Longtail-SWE tasks and repositories by language and difficulty.}
\label{tab:language-distribution}
\vskip 0.12in
\begin{tabular}{@{}lrrrrr@{}}
\toprule
Language & Tasks & Repos & Easy & Medium & Hard \\
\midrule
Clojure & 13 & 10 & 0 & 3  & 10 \\
Dart    & 14 & 11 & 1 & 3  & 10 \\
Elixir  & 23 & 13 & 0 & 7  & 16 \\
Erlang  & 12 & 7  & 2 & 4  & 6  \\
F\#     & 16 & 8  & 0 & 13 & 3  \\
Haskell & 14 & 8  & 1 & 5  & 8  \\
Julia   & 27 & 22 & 0 & 6  & 21 \\
Kotlin  & 12 & 12 & 0 & 1  & 11 \\
Lua     & 13 & 8  & 1 & 8  & 4  \\
OCaml   & 13 & 2  & 0 & 6  & 7  \\
R       & 14 & 9  & 1 & 11 & 2  \\
Scala   & 18 & 11 & 0 & 4  & 14 \\
Swift   & 13 & 8  & 1 & 2  & 10 \\
\midrule
Total   & 202 & 129 & 7 & 73 & 122 \\
\bottomrule
\end{tabular}
\end{table}

\subsection{Agent Model Configurations}
\label{app:prompts}

This section summarizes the model configurations and token usage for the agents used during task construction and repair (Sections~\ref{sec:method-construction} and~\ref{sec:method-repair}). Evaluation configurations are listed in Table~\ref{tab:overall-results} of the main paper; the complete agent prompts are provided in our public repository (Section~\ref{sec:method-release}).

The auditor and repairer use Kimi K3 with temperature 1, top\_p 1, high reasoning, and a 1M-token context window. The sequential judges use GPT 5.6 Luna (xhigh), GPT 5.6 Terra (high), and Kimi K3 (high). During description repair, solver attempts use Kimi K3 (high), GPT 5.6 Terra (high), and GPT 5.6 Luna (xhigh). After the complete statement and its addendum are frozen for the test-repair phase, solver attempts use GPT 5.6 Sol (high). Kimi K3 uses the same configuration across its auditor, repairer, judge, and construction-time solver roles.

The auditor, judge, and repairer average 2.8M, 0.7M, and 3.3M tokens per trajectory, respectively. Mean solver token usage is reported in Table~\ref{tab:overall-results} of the main paper.

\clearpage

\begin{lstlisting}[style=problemstatement,caption={Problem statement before repair for the \texttt{FlexiSummary} merge task.},label={lst:flexisummary-before}]
merge on FlexiSummary
If two `FlexiSummary`s have the same iter_indices and chain_indices, but different stat indices, we should be able to merge them.

In principle one may want to merge along different axes, but I think this is the main use case. For example, you can do

```julia
chain = ...

merge(mean(chain), hdi(chain; prob=0.95, split_interval=true))
```

and that would easily replicate the behaviour of tidybayes' (for example) `mean_hdi`.
\end{lstlisting}

\begin{lstlisting}[style=problemstatement,caption={Problem statement after repair for the \texttt{FlexiSummary} merge task, which shows additional requirements appended to the original problem statement.},label={lst:flexisummary-after}]
<THE ORIGINAL PROBLEM STATEMENT VERBATIM>

### Additional contract requirements

## Merging `FlexiSummary` objects

`Base.merge` is extended to `FlexiSummary`, with the following observable behavior:

- Two summaries can be merged provided that, for every collapsed dimension axis that both summaries retain, their iteration indices are equal and their chain indices are equal. If either the retained iteration indices or the retained chain indices differ, `merge` throws a `DimensionMismatch`. A summary collapsed along an axis has no indices for that axis, so such a collapsed axis places no equality constraint on the other summary for that axis.
- The merged summary contains the union of the keys of the input summaries and the union of their statistic names.
- Where a key is present in both summaries with the same statistic, the value from the later (rightmost) summary takes priority. Likewise, when a statistic name appears in both summaries, the later summary's values for that statistic take priority.
- When a key/statistic combination is present in only one of the input summaries, indexing the merged summary at that combination for the summary that lacks it yields `missing`.
- If both summaries have a statistic dimension marked as dropped (e.g. each summary was produced with a single statistic whose statistic dimension was dropped), then: if both summaries contain exactly the same statistic name, the merged summary also has its statistic dimension dropped; if the statistic names differ, the merged summary retains its statistic dimension over the combined statistic names.
- For a summary collapsed over iterations, the merged result likewise has `iter_indices(merged) === nothing`; for a summary collapsed over chains, `chain_indices(merged) === nothing`. Retained indices are preserved in the merged result.
- When the input summaries have different key types (for example `FlexiSummary{Symbol}` and `FlexiSummary{String}`), `merge` emits a warning mentioning "different key types" and returns a summary whose keys are promoted so that all original keys (with their original types) can still be accessed on the result.
- `merge` accepts three or more summaries, combining them as by successive pairwise merges.
\end{lstlisting}

\end{document}